\documentclass[jkps,preprint,fleqn,showpacs,showkeys]{revtex4}
\usepackage{graphicx}
\usepackage{amssymb}
\usepackage{amsmath}
\usepackage{epstopdf}
\usepackage{bm}
\usepackage{float}
\begin{document}
\setcounter{page}{1}
\title[]{Generalized Analytical Solutions and Synchronization Dynamics of Coupled Simple Nonlinear Electronic Circuits}
\author{G. Sivaganesh}
\email{sivaganesh.nld@gmail.com}

\affiliation{Department of Physics, Alagappa Chettiar College of Engineering $\&$ Technology, Karaikudi, Tamilnadu-630 004, India}
\author{A. Arulgnanam}
\affiliation{Department of Physics, St. John's College, Palayamkottai, Tamilnadu-627 002, India}

\date[]{Received 10 November 2016}

\begin{abstract}

In this paper we present a generalized analytical solution to the generalized state equations of coupled second-order non-autonomous circuit systems. The analytical solutions thus obtained are used to study the synchronization dynamics of two different types of circuit systems, differing only by their constituting nonlinear element. The synchronization dynamics of the coupled systems are studied through two-parameter bifurcation diagrams, phase portraits and time-series plots obtained from the explicit analytical solutions. The mechanism of synchronization is realized through the bifurcation of the eigenvalues as functions of the control parameter in each of the coupled piecewise linear regions of the drive and response systems. The stability of the synchronized state for coupled identical chaotic states are studeid through the {\emph{master stability function}}. Further, {\emph{conditional Lyapunov exponents}} and {\emph{Kaplan-Yorke dimension}} are obtained to confirm the synchronized states of both coupled identical and non-identical chaotic states. The synchronization dynamics of coupled chaotic systems studied through two-parameter bifurcation diagrams obtained from explicit analytical solutions is reported in the literature for the first time.
\end{abstract}



\maketitle

\section{Introduction}
\label{sec:1}

Synchronization of chaotic systems gained acceleration after the {\emph{Master-Slave}} concept introduced by Pecora and Carroll \cite{Pecora1990}. Since then, numerous chaotic systems and electronic circuits have been studied for their synchronization dynamics \cite{Chua1992,Chua1993,Murali1993,Murali1994a,Murali1995,Pecora1997,Murali1994}. Many natural systems have also been identified to exhibit chaos synchronization in their dynamics. Several synchronization phenomena such as generalized, complete, phase and lag synchronization were identified in identical and non-identical chaotic systems and were studied both experimentally and numerically. A detailed numerical study on the different types of synchronization phenomena has been presented by Boccaletti {\emph{et al.}} \cite{Boccaletti2002}. \\
Chaos synchronization has been observed in a variety of nonlinear electronic circuits because of their potential applications in secure transmission of signals \cite{Ogorzalek1993,Lakshmanan1994}. With the observation of a chaotic attractor in the {\emph{Chua's circuit}} \cite{Matsumoto1984}, nonlinear electronic circuits revealed the existence of chaotic attractors in circuits. The implementaion of the {\emph{Chua's diode}} using operational amplifiers \cite{Kennedy1992} enabled researchers to use the nonlinear element in a handful number of chaotic electronic circuits \cite{Murali1994,Venkatesan1999,Thamilmaran2000,Thamilmaran2001}. Further, nonlinear elements with different {\emph{v-i}} characteristics have been implemented using operational amplifiers \cite{Lacy1996,Arulgnanam2009,Bao2016}. Nonlinear circuits differ from each other mainly through through the the constitution of the nonlinear element that is present in the circuit. Hence, thier chaotic dynamics is decided by the type of the nonlinear element. The dynamics of those circuits have been mostly analyzed mostly through numerical experiments. The observation of chaos in a second-order non-autonomous circuit, namely the {\emph{Murali-Lakshmanan-Chua (MLC)}} circuit, by Murali {\emph{et al.}}, paved way to easily understand the complexity underlying the chaotic attractors. This lower dimensional system claims greater importance as it possesses a rich variey of chaotic dynamics resembling its higher dimensional counter parts. A few number of second-order non-autonomous chaotic systems with different circuit parameters and nonlinear elements exhibiting chaos were studied. Because of the mathematical simplicity of second-order, piecewise linear chaotic systems, analytical solutions to their normalized state equations has been obtained for a good number of chaotic systems \cite{Lakshmanan1995,Lakshmanan1995a,Thamilmaran2001,Lakshmanan2003,Thamilmaran2005,Arulgnanam2009,Sivaganesh2014,Arulgnanam2015}. However explicit analytical solutions explaining the synchronization dynamics of coupled systems are rare in literature. Explicit analytical solutions explaining the complete synchronization phenomenon in coupled second-order non-autonomous circuits have were studied recently \cite{Sivaganesh2015,Venkatesh2016,Sivaganesh2016a}.\\

We present in this paper, an analytical study on the synchronization dynamics observed in coupled simple second-order dissipative systems. The synchronization dynamics of two types of second-order non-autonomous circuits are studied. The two circuits differ only by the nonlinear element contained in the circuit. The first circuit consists of forced parallel LCR circuit with a {\emph{Chua's diode}} as the nonlinear element. The synchronization dynamics of this circuit is studied for coupled identical and non-identical chaotic attractors. The second circuit has a forecd parallel LCR circuit with the {\emph{simplified nonlinear element}} as the nonlinear element. The synchronization dynamics of this circuit is also studied for coupled identical and non-identical chaotic attractors. As both the circuits have nonlinear elements  with piecewise linear charactersistics the solutions to their normalized state equations can be generalized.  Further, generalized analytical solutions to the normalized state equations of the unidirectionally coupled circuits have been obtained. The analytical solutions thus obtained have been used to generate two-parameter bifurcation diagrams to indicate the complete synchronization dynamics of the coupled systems. The stability of synchronization of coupled identical chaotic systems is studied using the {\emph{Master Stability Function (MSF)}} \cite{Pecora1998,Liang2009}. The MSFs for coupled identical {\emph{variant of MLC}} and series LCR circuit with a {\emph{simplified nonlinear element}} circuits for coupling of different state variables has been studied \cite{Sivaganesh2016}. A simple analytical technique is introduced to derive explicit analytical solutions to the normalized state equations of the coupled system. This is achieved by mere coupling of similar picewise linear regions of the drive and response systems. A linear stability analysis on the eigenvalues of the coupled system reveals the mechanism of complete synchronization. The onset of synchronization identified through stability analysis is confirmed by the numerical study of the MSF. The synchronization dynamics studied through the analytical solutions has been substantiated through numerical studies of the {\emph{conditional Lyapunov exponents}} and the {\emph{Kaplan-Yorke dimension}} $(D_{KY})$ in each case. Because the sudden rise or fall of $(D_{KY})$ is a measure of unsynchronization or synchronization, it is studied for each case of the coupled systems.

This paper is divided into five sections. In Section \ref{sec:2} we present the generalized circuit equations and generalized analytical solutions to the circuit equations of the forced parallel LCR circuit with a nonlinear element. The analytical technique to find explicit analytical solutions to the normalized state equations of the coupled system is presented in Section \ref{sec:3}. In Section \ref{sec:4} we present the explain the synchronization dynamics observed in coupled {\emph{variant of MLC circuits}} using the analytical solutions obtained from Section \ref{sec:3}. The synchronization dynamics observed in coupled series LCR circuits with {\emph{a simplified nonlinear element}} is explained in Section \ref{sec:5}.

\section{Circuit Equations}
\label{sec:2}

\begin{figure}[H]
\begin{center}
\resizebox{0.66\textwidth}{!}{%
  \includegraphics{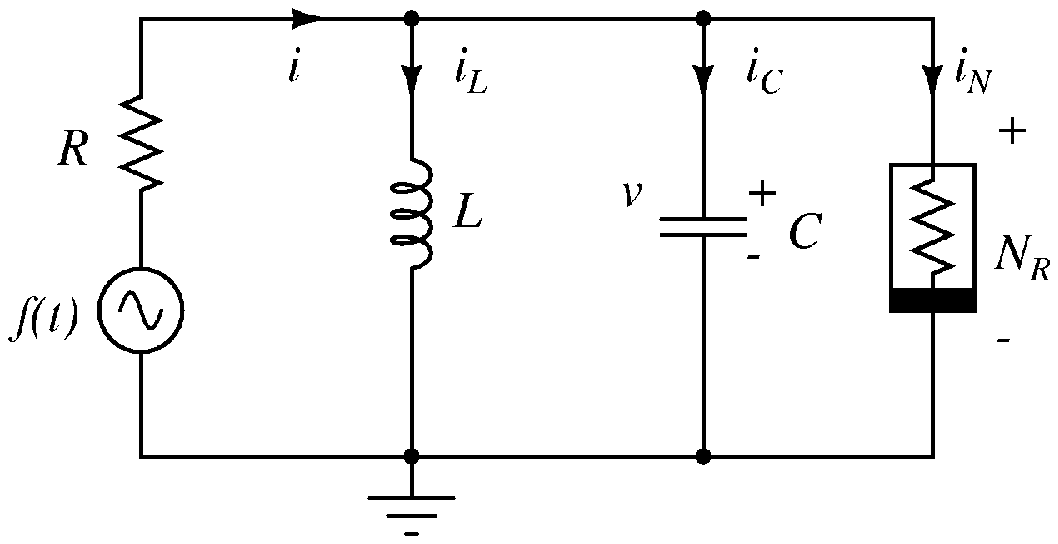}
}
\caption{(a) Schematic circuit realization of the sinusoidally forced parallel LCR circuit with the nonlinear element $N_R$ connected parallel to the capacitor.}
\label{fig:1}       
\end{center}
\end{figure}

\begin{figure}
\resizebox{0.66\textwidth}{!}{%
  \includegraphics{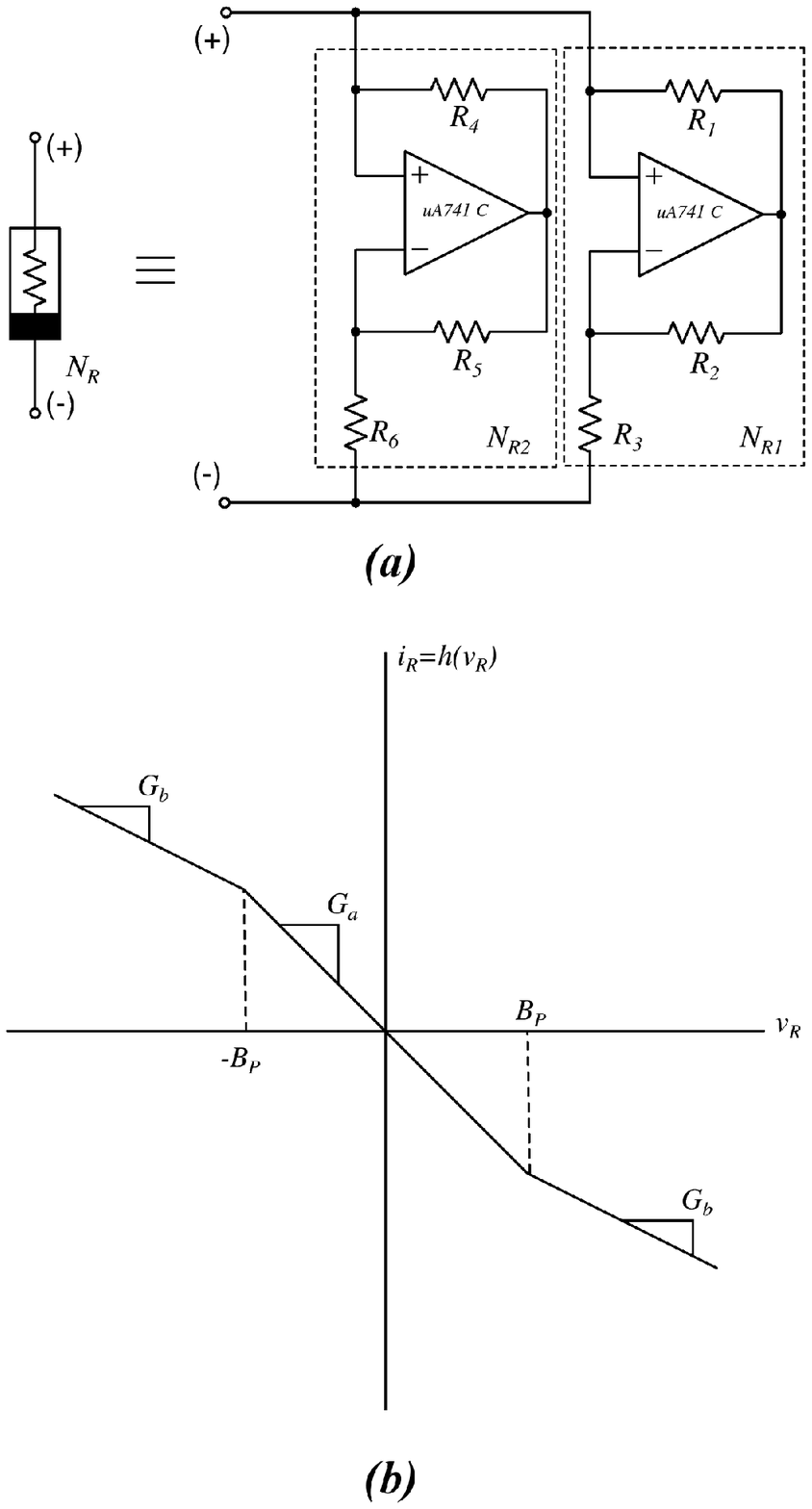}
}
\caption{(a) Schematic realization of the {\emph{Chua's diode}} constructed from two operational amplifiers and six linear resistors and its (b) $(v-i)$ characteristics with one negative inner slope $G_{a} = -0.76$ mS, two negative outer slopes $G_{b} = -0.41$ mS and the break points $B_{p} = \pm~1.0$ V, respectively.}
\label{fig:2}       
\end{figure}

\begin{figure}
\resizebox{0.66\textwidth}{!}{%
  \includegraphics{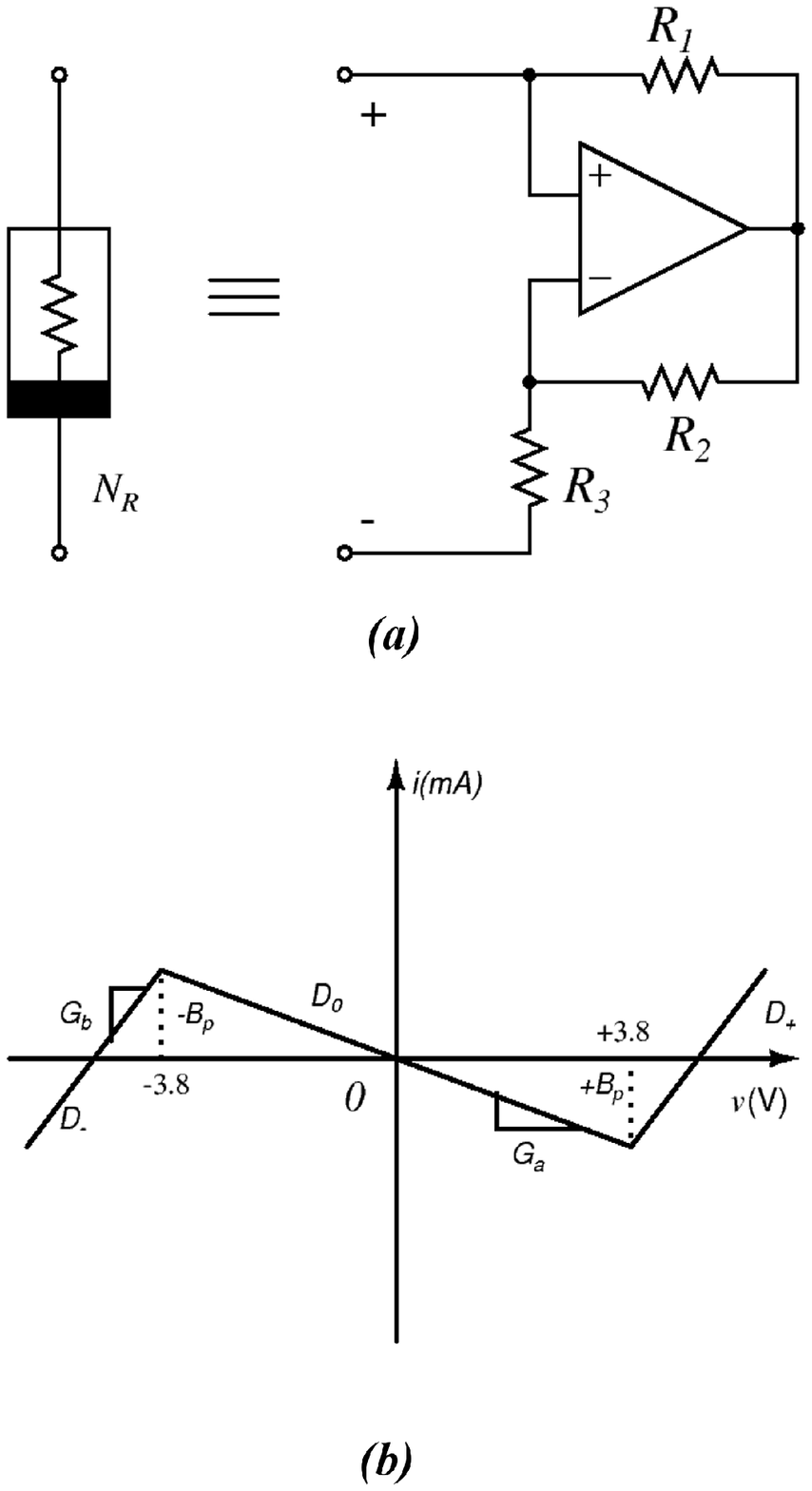}
}
\caption{(a) Schematic realization of the {\emph{simplified nonlinear element}} constructed using one operational amplifier and three linear resistors and its (b) $(v-i)$ characteristics with a negative inner slope $G_{a} = -0.56$ mS, two positive outer slopes $G_{b} = +2.5$ mS and the break points $B_{p} = \pm~3.8$ V, respectively.}
\label{fig:3}       
\end{figure}

A sinusoidally forced parallel LCR circuit with a nonlinear element $N_R$ connected parallel to the capacitor, exhibiting chaotic dynaimics is as shown in Fig. \ref{fig:1}. The nonlinear element $N_R$ shown in Fig. \ref{fig:1} determines the chaotic dynamics of the circuit for a proper choice of the other circuit parameters. The nonlinear elements considered for the present study were voltage-controlled and piecewise-linear. We consider two types of nonlinear elements for $N_R$ namely, the {\emph{Chua's diode}} and the {\emph{simplified nonlinear element}}. The {\emph{Chua's diode}} is a piecewise-linear nonlinear element with three negative slope regions has been constructed using two operational amplifiers and six linear resistors as shown in Fig. \ref{fig:2}(a). The $(v-i)$ charcateristics of the {\emph{Chua's diode}} with two negative outer slopes $(G_b)$ and one negative inner slope $(G_a)$ is as shown in Fig. \ref{fig:2}(b). A parallel LCR circuit with the {\emph{Chua's diode}} as the nonlinear element has been called as the {\emph{Variant of Murali-Lakshmanan-Chua}} circuit was introduced by Thamilmaran \cite{Thamilmaran2000}. The torus breakdown, antimonocity, intermittency, period doubling and the reverse period doubling routes to chaos exhibited by the $MLCV$ circuit has been extensively studied \cite{Thamilmaran2000,Thamilmaran2001}. Further, an explicit analytical solution to the state variables of the normalized circuit equations has been presented  \cite{Thamilmaran2001}. A {\emph{simplified nonlinear element}} is a kind of the modified {\emph{Chua's diode}} was introduced by Arulgnanam \cite{Arulgnanam2009} is as shown in Fig. \ref{fig:3}(a). It has been constructed with one operational amplifier and three linear resistors and making it the simplest nonlinear element constructed from a least number of circuit elements. The $(v-i)$ characteristics of the {\emph{simplified nonlinear element}} is also three segmented but with one inner negative slope $(G_a)$ and two positive outer slope regions $(G_b)$ as shown in Fig. \ref{fig:3}(b). It has been evidenced that the maximal Lyapunov exponents observed in circuits with {\emph{simplified nonlinear element}} is relatively higher than that with the {\emph{Chua's diode}} as teh nonlinear element \cite{Arulgnanam2009}. In both the nonlinear elements, the central region $(D_0)$ is always negative and the outer regions $(D_{\pm})$ are positive for the {\emph{simplified nonlinear element}} while they are negative for the {\emph{Chua's diode}}. However, the circuit equations governing the {\emph{forced parallel LCR circuit}} with either of the nonlinear element remains the same. The equations of the circuit shown in Fig. \ref{fig:1} in terms of the voltage $(v)$ across the capacitor and current $(i_L)$ flowing through the inductor, using {\emph{Kirchoff's laws}} is given as 
\begin{subequations}
\begin{eqnarray} 
C {dv \over dt } & = & {1\over R} (F sin( \Omega t)- v)- i_L - g(v), \\
L {di_L \over dt } & = & v,
\vspace{-1.5cm}
\end{eqnarray}
\end{subequations}
where $g(v)$ is the mathematical form of the voltage-controlled, piecewise-linear resistor given by
\begin{equation}
g(v) = G_b v + 0.5(G_b - G_a)[|v+B_p|-|v-B_p|]
\end{equation}
The terms $f$, $\omega$ represent the amplitude and frequency of the external sinusoidal voltage source respectively. Using the rescaling parameters $x = v/B_p$, $y = (i_L / G B_p)$,  $ \beta = (C/LG^2)$, $ a  = G_a/G$,  $ b = G_b/G$, $f = (F \beta/B_p)$, $\omega = (\Omega C/G)$ and  $ G = 1/R$, the normalized state equations of the circuit can be written as
\vspace{0.1cm}
\begin{subequations}
\begin{eqnarray}
\dot x & = & f sin(\theta) - x - y - g(x), \\
\dot y & = & \beta x , \\
\dot \theta & = & \omega,
\end{eqnarray}
\label{eqn:3}
\end{subequations}
The function $g(x)$ can be written in the piecewise linear form as 
\begin{equation}
g(x) =
\begin{cases}
bx+(a-b) & \text{if $x \ge 1$}\\
ax & \text{if $|x|\le 1$}\\
bx-(a-b) & \text{if $x \le -1$}
\end{cases}
\end{equation}
The circuit parameters are so chosen to obtain chaotic behavior in its dynamics. The circuit parameters take different values when different nonlinear elements are used. Based on the type of nonlinear element used, the circuit exhibits two different types of chaotic attractors. First we summarize the explicit analytical solutions obtained in general for the normalized state equations given in Eq. \ref{eqn:3}. The analytical solutions are obtained for each of the piecewise-linear regions of the nonlinear element. In both the nonlinear elements, the central region has been taken as $D_0$ and the outer regions in the positive and negative voltage regime as $D_+$ and $D_-$, respectively. 

\subsubsection{\ $Region: D_0$}

In the central region we have $g(x)=ax$ with $(0,0)$ as the fixed point. The roots ${m_{1,2}} =  \frac{-(A) \pm \sqrt{(A^{2}-4B)}} {2}$ may be either two real values or a pair of complex conjugates. When the roots are a pair of complex conjugates, the state variables $y(t)$ and $x(t)$ are
\begin{subequations}
\begin{eqnarray}
y(t) &=& e^ {ut}(C_1 cos vt + C_2 sin vt) +E_1 sin(\omega_1 t)+ E_2 cos(\omega_1 t),\\
x(t) &=& \frac{1}{\beta}(\dot{y}), 
\end{eqnarray}
\label{eqn:5}
\end{subequations}
where, $u=\frac{-A}{2}$ and $v=\frac{\sqrt(4B-A^{2})}{2}$. When the roots are two real values then the state variables are
\begin{subequations}
\begin{eqnarray}
y(t) &=& C_1 e^ {m_1 t} + C_2 e^ {m_2 t} + E_1 + E_2 \sin \omega_1 t + E_3 \cos \omega_1 t,\\
x(t) &=& \frac{1}{\beta}(\dot{y}), 
\end{eqnarray}
\label{eqn:6}
\end{subequations}

\subsubsection{\ $Region: D_{\pm1}$}

In outer regions $D_{\pm 1}$, we have $g(x)=bx \pm (a-b)$ with $k_1=0$, $k_2={\pm}(a-b)$ as the fixed point. The state variables $y(t)$ and $x(t)$ in this region when the roots ${m_{3,4}} =  \frac{-(A) \pm \sqrt{(A^{2}-4B)}} {2}$ are a pair of complex conjugates are given as 

\begin{subequations}
\begin{eqnarray}
y(t) &=& e^ {ut}(C_3 cos vt + C_4 sin vt) +E_3 sin(\omega_1 t)+ E_4 cos(\omega_1 t) {\pm} \Delta,\\
x(t) &=& \frac{1}{\beta}(\dot{y}), 
\end{eqnarray}
\label{eqn:7}
\end{subequations}
When the roots are two real values then the state variables are
\begin{subequations}
\begin{eqnarray}
y(t) &=& C_3 e^ {m_3 t} + C_4 e^ {m_4 t} + E_3 \sin \omega_1 t + E_4 \cos \omega_1 t {\pm} \Delta,\\
x(t) &=& \frac{1}{\beta}(\dot{y}), 
\end{eqnarray}
\label{eqn:8}
\end{subequations}
where $\Delta = (b-a)$ and $+\Delta$, $-\Delta$ corresponds to $D_{+1}$ and $D_{-1}$ regions respectively. 

The circuit explained above acting as the drive system can be unidirectionally coupled to another circuit with the same nonlinear element acting as the response system. The drive and the response system operating with different set of initial conditions have been coupled together by a linear resistor and a buffer. The buffer acts as a signal driving element which isolates the drive system variables being affected by the response system. The schematic diagram of the coupled circuits is as shown in Fig. \ref{fig:2}. The normalized state equations of the response system can be written as
\begin{subequations}
\begin{eqnarray}
\dot {x^{'}} & = & f_2 sin(\omega_2 t) - x^{'} - y^{'} - g(x^{'})+\epsilon(x-x^{'}),\\
\dot {y^{'}} & = & \beta x^{'},\\
\dot {\theta^{'}} & = & \omega_2.
\end{eqnarray}
\label{eqn:9}
\end{subequations}
and the piecewise linear function $g(x^{'})$ is
\begin{equation}
g(x^{'}) =
\begin{cases}
bx^{'}+(a-b) & \text{if $x^{'}\ge 1$}\\
ax^{'} & \text{if $|x^{'}|\le 1$}\\
bx^{'}-(a-b) & \text{if $x^{'}\le -1$}
\end{cases}       
\end{equation}
where, $f_2 = (F_2 \beta/B_p)$, $\omega_2 = (\Omega_2 C/G)$  and $\epsilon = (R/R_c)$ is the coupling parameter. The state variables of the response system have been represented as $x^{'},y^{'}$. The amplitude and frequency of the external forcing term of the drive and response systems have been taken as $f_1, \omega_1$ and $f_2, \omega_2$, respectively.

\section{Explicit Analytical Solutions}
\label{sec:3}

In this section, we present explicit analytical solutions for the normalized circuit equations of the response system given in Eq. \ref{eqn:9}. From the state equations of the response system we observe that the dynamics of the response is influenced by the drive through the coupling parameter. Because the circuit equations are piecewise linear, each picewise linear region of the two sytems could be coupled together to get a new set of equations which could be solved for each region. The solution obtained for each piecewise linear region could be matched across the boundaries to study the dynamics of the coupled system. The new set of equations obtained from Eqs. \ref{eqn:3} and \ref{eqn:9} are
\begin{subequations}
\begin{eqnarray}
\dot {x^{*}} &=& f_1 sin(\omega_1 t) - f_2 sin(\omega_2 t) - x^{*} - y^{*} - g(x^{*})-\epsilon x^{*},\\
\dot {y^{*}} &=& \beta x^{*}, 
\end{eqnarray}
\label{eqn:11}
\end{subequations}
where $x^{*}$=$(x-x^{'})$, $y^{*}$=$(y-y^{'})$ and $g(x^{*}) = g(x) - g(x^{'})$ takes the values $a{x^{*}}$ or $b{x^{*}}$ depending upon the region of operation of the drive and response system. From the new set of state variables $x^{*}(t),~y^{*}(t)$, the state variables of the response system  $x^{'}(t),~y^{'}(t)$ could be written as
\begin{subequations}
\begin{eqnarray}
x^{'} &=& x -  x^{*}, \\
y^{'} &=& y -  y^{*}.
\end{eqnarray}
\label{eqn:12}
\end{subequations}


One can easily establish that a unique equilibrium point $ (x^{*}_0,y^{*}_0)$ exists for Eq. \ref{eqn:11} in each of the following three subsets
\begin{equation}
\left.
\begin{aligned}
D^{*}_{+1} & =  \{ (x^{*},y^{*})| x^{*} > 1 \} P^{*}_+ = (0,0),\\
D^{*}_0 & =  \{ (x^{*},y^{*})|| x^{*} | < 1 \}| O^{*} = (0,0),\\
D^{*}_{-1} & =  \{ (x^{*},y^{*})|x^{*} < -1 \}| P^{*}_- = (0,0),\\
\end{aligned}
\right\}
\quad\text{}
\label{eqn:13}
\end{equation}
Naturally these fixed points can be observed depending upon the initial conditions $x^{*}_0$ and $y^{*}_0$ of Eqs. \ref{eqn:11} which are inturn obtained from the initial conditions of the drive and response systems as, $x^{*}_0 = x_0 - x^{'}_0$ and $y^{*}_0 = y_0 - y^{'}_0$. Because the origin $(0,0)$ becomes the fixed points for all the three piecewise-linear regions $D^{*}_0$ and $D^{*}_{\pm 1}$, the fixed points of the drive and response system remains same under the uncoupled state. The stability of the fixed points given in Eq. \ref{eqn:13} can be calculated from the stability matrices. In the first case, $g(x)$ and $g(x^{'})$ take the values  $a{x}$ and $a{x^{'}}$ respectively, corresponding to the central region in the $(v-i)$ characteristics of the nonlinear element, which has been taken as the $D^{*}_0$ region of the difference system. In the $D^{*}_0$ region the stability determining eigenvalues are calculated from the stability matrix
\begin{equation}
J^{*}_{0} =
\begin{pmatrix}
-(a+\epsilon+1) &&& -1 \\
\beta &&& 0 \\
\end{pmatrix},
\label{eqn:14}
\end{equation}

In the second case, $g(x)$ and $g(x^{'})$ take the values  $bx \pm (a-b)$ and $bx^{'} \pm (a-b)$ respectively, corresponding to the outer regions in the $v-i$ characteristics of the nonlinear element, which has been taken as the $D^{*}_{\pm1}$ regions of the difference system.
In the $D^{*}_{\pm1}$ regions, the stability determining eigen values are calculated from the stability matrix
\begin{equation}
J^{*}_{\pm} =
\begin{pmatrix}
-(b+\epsilon+1) &&& -1 \\
\beta &&& 0 \\
\end{pmatrix}
\label{eqn:15}
\end{equation}
The eigenvalues of the difference system in both the regions are thus determined by the strength of the coupling parameter. 


 When the coupling paramter $\epsilon=0$, the coupled systems become independent of each other. Hence the drive and the response systems given by Eqs. \ref{eqn:3} and \ref{eqn:9} have the same solution for their state variables in all the three piecewise linear regions. 


Now we present explicit analytical solutions for the dynamics of the response system for coupling strengths $\epsilon > 0$. An explicit analytical solution to the dynamics of the response system could be obtained by finding a solution to the normalized state variables of the difference system given by Eq. \ref{eqn:11}. The solution of those equations are, $ [x^{*} (t; t_0, x^{*}_0, y^{*}_0), ~y^{*}(t; t_0, x^{*}_0, y^{*}_0)]^T$ for which the initial conditions are written as $ (t, x^{*}, y^{*}) $ $ = (t_0, x^{*}_0, y^{*}_0) $. From the solution $x^{*}(t)$ and $y^{*}(t)$ thus obtained the state variables $x^{'}(t)$ and $y^{'}(t)$  can be found using Eqs. \ref{eqn:12}. Since Eq. \ref{eqn:11} is piecewise linear, the solution to each of the three regions can be obtained explicitly. 

\subsubsection{ \bf $Region: D^{*}_0$}

In this region $g(x)$ and $g(x^{'})$ takes the values  $a{x}$ and $a{x^{'}}$ respectively. Hence the normalized equations obtained from Eqs. \ref{eqn:11} are
\begin{subequations}
\begin{eqnarray}
\dot {x^{*}} & = & - (a+\epsilon+1)x^{*} - y^{*}+f_1 sin(\omega_1 t) - f_2 sin(\omega_2 t), \\
\dot {y^{*}} & = & \beta x^{*},
\end{eqnarray}
\label{eqn:16}
\end{subequations}
Differentiating Eq. (\ref{eqn:16}b) with respect to time and using Eqs. (\ref{eqn:16}a, \ref{eqn:16}b) in the resultant equation, we obtain
\begin{equation}
{\ddot y^{*}} + {A \dot y^{*}} + By^{*} = f_1 sin(\omega_1 t) - f_2 sin(\omega_2 t),
\label{eqn:17}
\end{equation}
where, $ A =  a + \epsilon +1$ and  $B = \beta$.
The roots of the Eq. \ref{eqn:17} is given by ${m_{1,2}} =  \frac{-(A) \pm \sqrt{(A^{2}-4B)}} {2}$. Since the roots ${m_{1,2}}$ depends on the coupling parameter, the orientation of the trajectories around the fixed point changes as the coupling parameter is varied. 
\subsection*{case: a}

When $(A^{2} < 4B)$, the roots $m_1$ and $m_2$ are a pair of complex conjugates given as $m_{1,2} = u \pm iv$, with $u=\frac{-A}{2}$ and $v=\frac{\sqrt(4B-A^{2})}{2}$. Using the method of undetermined coefficients, the general solution to Eq. \ref{eqn:17} can be written as
\begin{equation}
y^{*}(t) =  e^ {ut} (C_1 cosvt+ C_2 sinvt)+ E_1 sin \omega_1 t + E_2 cos \omega_1 t + E_3 sin \omega_2 t + E_4 cos \omega_2 t,
\label{eqn:18}
\end{equation}
where $C_1$ and $C_2$ are integration constants and
\begin{eqnarray}
E_1  &=&  \frac {f_1 \beta (B-{\omega_1}^2)}{A^2 {\omega_1} ^2 + (B-{\omega_1} ^2)^2}, \nonumber \\
E_2  &=&  \frac {-A \omega_1 f_1 \beta}{A^2 {\omega_1} ^2 + (B-{\omega_1} ^2)^2}, \nonumber \\
E_3  &=&   \frac {-f_2 \beta (B-{\omega_2}^2)}{A^2 {\omega_2} ^2 + (B-{\omega_2} ^2)^2}, \nonumber \\
E_4  &=&  \frac {A \omega_2 f_2 \beta}{A^2 {\omega_2} ^2 + (B-{\omega_2} ^2)^2}, \nonumber
\end{eqnarray}
Differentiating Eq. \ref{eqn:18} and using it in Eq. (\ref{eqn:16}b) we get
\begin{equation}
x^{*}(t) = \frac{1}{\beta}(\dot{y^{*}}),
\label{eqn:19}
\end{equation}
The constants $C_1$ and $C_2$ in the above equations can be evaluated by solving both Eqs. \ref{eqn:18} and \ref{eqn:19} for $C_1$ and $C_2$ at a suitable initial instant $t_0$, with $x^{*}_0$ and $y^{*}_0$ as initial conditions at time $t=t_0$, provided the trajectory of the dynamical system just enters the region $D^{*}_0$ at time $t_0$. The constants $C_1$ and $C_2$ thus obtained are \\
\begin{eqnarray}
C_1 =  \frac{e^ {- u t_0}} {v} \{(v { y^{*}_0} cos vt_0- (\beta { x^{*}_0}-u { x^{*}_0}) sin vt_0)+((\omega_1 E_1 - u E_2) sinvt_0 - vE_2 cosvt_0)cos \omega_1 t_0 \nonumber \\
             - ((\omega_1 E_2 +u E_1) sinvt_0+v E_1 cosvt_0) sin \omega_1 t_0 + ((\omega_2 E_3 - u E_4) sinvt_0 - vE_4 cosvt_0)cos \omega_2 t_0  \nonumber \\
	  - ((\omega_2 E_4 + u E_3) sinvt_0 + vE_3 cosvt_0) sin \omega_2 t_0 \},  \nonumber \\
C_2 =  \frac{e^ {- u t_0}} {v} \{((\beta { x^{*}_0}-u { x^{*}_0}) cos vt_0+v { y^{*}_0} sin vt_0)-((\omega_1 E_1 - u E_2) cos vt_0 + v E_2 sin vt_0)cos \omega_1 t_0 \nonumber \\
             + ((\omega_1 E_2 +u E_1) cos vt_0 - v E_1 sin vt_0) sin \omega_1 t_0 - ((\omega_2 E_3 - u E_4) cos vt_0 + vE_4 sin vt_0)cos \omega_2 t_0  \nonumber \\
	  + ((\omega_2 E_4 + u E_3) cos vt_0 - vE_3 sin vt_0) sin \omega_2 t_0 \}, \nonumber 
\end{eqnarray}
From the results of $y^{*}(t),~x^{*}(t)$ obtained from Eqs. \ref{eqn:18}, \ref{eqn:19} and $y(t),~x(t)$ obtained from Eqs. \ref{eqn:5}(a), \ref{eqn:5}(b), $x^{'}(t)$ and $y^{'}(t)$ can be obtained from Eqs. \ref{eqn:12}.

\subsubsection*{case: b}

When $(A^{2} > 4B)$, the roots $m_1$ and $m_2$ are real and distinct.
The general solution to Eq. \ref{eqn:17} can be written as
\begin{equation}
y^{*}(t) = C_1 e^ {m_1 t} + C_2 e^ {m_2 t} + E_1 sin(\omega_1 t) + E_2 cos(\omega_1 t) + E_3 sin(\omega_2 t) + E_4 cos(\omega_2 t),
\label{eqn:20}
\end{equation}
where $C_1$ and $C_2$ are the integration constants and the constants $E_1,E_2,E_3,E_4$ are the same as in {\emph{case: a}}.
Differentiating Eq .\ref{eqn:20} and using it in Eq.(\ref{eqn:16}b) we get
\begin{equation}
x^{*}(t) = \frac{1}{\beta}(\dot{y^{*}}),
\label{eqn:21}
\end{equation}
The constants $C_1$ and $C_2$ are
\begin{eqnarray}
C_1 =  \frac{e^ {- m_2 t_0}} {m_2 - m_1} \{ (\beta{ x^{*}_0} - m_1{ y^{*}_0}) + ( m_1 E_2 - \omega_1 E_1) cos \omega_1 t_0 + (\omega_1 E_2 + m_1 E_1) sin \omega_1 t_0 \nonumber \\
            +(m_2 E_4 - \omega_2 E_3) cos \omega_2 t_0 + (\omega_2 E_4 + m_2 E_3 ) sin \omega_2 t_0 \}, \nonumber \\
C_2 =  \frac{e^ {- m_1 t_0}} {m_1 - m_2} \{ (\beta{ x^{*}_0} - m_2{ y^{*}_0}) + ( m_2 E_2 - \omega_1 E_1) cos \omega_1 t_0 + (\omega_1 E_2 + m_2 E_1) sin \omega_1 t_0 \nonumber \\
            +(m_1 E_4 - \omega_2 E_3) cos \omega_2 t_0 + (\omega_2 E_4 + m_1 E_3 ) sin \omega_2 t_0 \}, \nonumber
\end{eqnarray} 
From the results of $y^{*}(t),~x^{*}(t)$ obtained from Eqs. \ref{eqn:20}, \ref{eqn:21} and $y(t),~x(t)$ obtained from Eqs. \ref{eqn:6}(a), \ref{eqn:6}(b), $x^{'}(t)$ and $y^{'}(t)$ can be odtained from Eqs. \ref{eqn:12}.

\subsubsection{\bf $Region: D^{*}_{\pm1}$}

In this region $g(x)$ and $g(x^{'})$ takes the values  $b{x}\pm (a-b)$ and $b{x^{'}} \pm (a-b)$ respectively. Hence the normalized state equations obtained from Eqs. \ref{eqn:11} are
\begin{subequations}
\begin{eqnarray}
\dot {x^{*}} & = & - (b+\epsilon+1)x^{*} - y^{*}+f_1 sin(\omega_1 t) - f_2 sin(\omega_2 t), \\
\dot {y^{*}} & = & \beta x^{*},
\end{eqnarray}
\label{eqn:22}
\end{subequations}
Differentiating Eq. (\ref{eqn:22}b) with respect to time and using Eqs. (\ref{eqn:22}a, \ref{eqn:22}b) in the resultant equation, we obtain
\begin{equation}
{\ddot y^{*}} + {C \dot y^{*}} + Dy^{*} = f_1 sin(\omega_1 t) - f_2 sin(\omega_2 t),
\label{eqn:23}
\end{equation}
where, $ C =  b + \epsilon +1$ and  $D = \beta$.
The roots of the Eq. \ref{eqn:23} is given by ${m_{3,4}} =  \frac{-(C) \pm \sqrt{(C^{2}-4D)}} {2}$.

\subsubsection*{case: a}

When $(C^{2} < 4D)$, the roots $m_3$ and $m_4$ are a pair of complex congugates given as $m_{3,4} = u \pm iv$, with $u=\frac{-C}{2}$ and $v=\frac{\sqrt(4D-C^{2})}{2}$. Using the method of undetermined coefficients, the general solution to Eq. \ref{eqn:23} can be written as
\begin{equation}
y^{*}(t) =  e^ {ut} (C_3 cosvt+ C_4 sinvt)+ E_5 sin \omega_1 t + E_6 cos \omega_1 t + E_7 sin \omega_2 t + E_8 cos \omega_2 t,
\label{eqn:24}
\end{equation}
The constants $E_5,E_6, E_7, E_8$ are the same  as the contants $E_1, E_2, E_3, E_4$ in $case:a$ of $D^{*}_{0}$ region except that the constants $A,~B$ are replaced with $C,~D$ respectively. 
Differentiating Eq. \ref{eqn:24} and using it in Eq. (\ref{eqn:22}b) we get
\begin{equation}
x^{*}(t) = \frac{1}{\beta}(\dot{y^{*}}).
\label{eqn:25}
\end{equation}
The constants $C_3$ and $C_4$ are the same as $C_1$ and $C_2$ in $(case:a)$ of $D^{*}_{0}$ region except that the constants $E_1, E_2, E_3, E_4$ are replaced with the constants $E_5,E_6, E_7, E_8$ respectively.
From the results of $y^{*}(t),~x^{*}(t)$ obtained from  Eqs. \ref{eqn:24}, \ref{eqn:25} and $y(t),~x(t)$ obtained from  Eqs. \ref{eqn:7}(a), \ref{eqn:7}(b), $x^{'}(t)$ and $y^{'}(t)$ can be obtained from Eq. \ref{eqn:12}.

\subsubsection*{case: b}

When $(C^{2} > 4D)$, the roots $m_3$ and $m_4$ are real and distinct.
The general solution to Eq. \ref{eqn:23} can be written as
\begin{equation}
y^{*}(t) = C_3 e^ {m_3 t} + C_4 e^ {m_4 t} + E_5 sin(\omega_1 t) + E_6 cos(\omega_1 t) + E_7 sin(\omega_2 t) + E_8 cos(\omega_2 t),
\label{eqn:26}
\end{equation}
where $C_3$ and $C_4$ are the integration constants and the constants $E_5,E_6,E_7,E_8$ are the same as in {\emph{(case: a)}}.
Differentiating Eq. \ref{eqn:26} and using it in Eq. (\ref{eqn:22}b) we get
\begin{equation}
x^{*}(t) = \frac{1}{\beta}(\dot{y^{*}}).
\label{eqn:27}
\end{equation}
The constants $C_3$ and $C_4$ are the same as $C_1$ and $C_2$ in {\emph{(case:a)}} of $D^{*}_{0}$ region except that the constants $E_1, E_2, E_3, E_4$ are replaced with the constants $E_5,E_6, E_7, E_8$ respectively. From the results of $y^{*}(t),~x^{*}(t)$ obtained from Eqs. \ref{eqn:26}, \ref{eqn:27} and $y(t),~x(t)$ obtained from Eqs. \ref{eqn:8}(a), \ref{eqn:8}(b), $x^{'}(t)$ and $y^{'}(t)$ can be obtained from Eq. \ref{eqn:12}.\\

Now let us briefly explain how the solution can be generated in the $(x^{'}-y^{'})$ phase space. The analytical solutions obtained above can be used to simulate the trajectories of the state variables $y^{*}(t)$ and $x^{*}(t)$. With the {\emph{time (t)}} being considered as the independent variable, the state variables evolves within each piecewise-linear region depending upon its initial values. If we start with the initial conditions $x^{*}(t=0) = x^{*}_0,~y^{*}(t=0) = y^{*}_0$ in the $D^{*}_0$ region at time $t=0$, the arbitrary constants $C_1$ and $C_2$ get fixed. Thus $x^{*}(t)$ evolves as given by Eq. \ref{eqn:19} up to either $t=T_1$, when $x^{*}(T_1)=1$ or $t = T^{'}_1$ when $x^{*}(T^{'}_1) = -1$. The next region of operation $(D^{*}_{+1}$ or $D^{*}_{-1})$ thus depends upon the value of $x^{*}$ in the $D^{*}_{0}$ region at that instant of time. As the trajectory enters into the next region of interest, the arbitrary constant corresponding to that region could be evaluated, with the initial conditions to that region being either ($x^{*}_{0}(T_1),~y^{*}_{0}(T_1)$) or  ($x^{*}_{0}(T^{'}_1),~y^{*}_{0}(T^{'}_1)$). During each region of operation, the state variables of the response system evolves as $x^{'}(t) = x(t) - x^{*}(t)$ and $y^{'}(t) = y(t) - y^{*}(t)$, respectively. The procedure can be continued for each successive crossing. In this way, the explicit solutions can be obtained in each of the regions $D^{'}_0$, $D^{'}_{\pm1}$ of the response system. The solution obtained in each region has been matched across the boundaries and used to generate the dynamics of the response system.  \\ 

\section{Synchronization dynamics of the variant of MLC circuit}
\label{sec:4}

\begin{figure}
\resizebox{0.66\textwidth}{!}{%
  \includegraphics{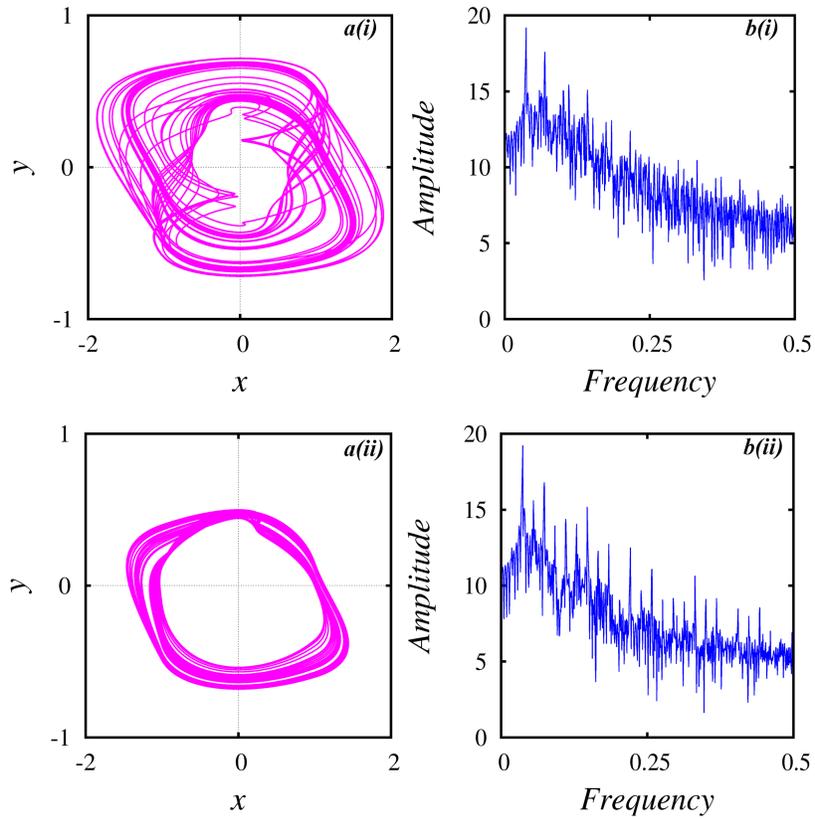}
}
\caption{Analytically obtained chaotic attractors and their power spectrum for the circuit with the {\emph{Chua's diode}} as the nonlinear element. (a) Left panel represent the chaotic attractors obtained at $f=0.39$ and $f=0.411$ and (b) right panels represent their corresponding power spectrum.}
\label{fig:5}       
\end{figure}

In this section we discuss the synchronization dynamics observed in the coupled system when the {\emph{Chua's diode}} has been used as the nonlinear element. A sinusoidally forced parallel LCR circuit with the {\emph{Chua's diode}} as the nonlinear element has been called as the {\emph{Variant of Murali-Lakshmanan-Chua}} circuit. The circuit exhibits chaotic dynamics for the chosen circuit parameter values circuit parameters $C=10.15$ nF, $L=445$ mH, $R=1475~\Omega$. The negative slopes of the inner, outer regions and the breakpoints in the $(v-i)$ characteristic curve of the piecewise-linear element shown in Fig. \ref{fig:1}(b) are given as $G_{a} = -0.76$ mS, $G_{b} = -0.41$ mS and $B_{p} = \pm~1.0$ V, respectively. The rescaled circuit parameters take the values as $\beta = 0.05,~a=-1.121,~b=-0.6047$. The frequency of the external periodic force has been fixed at $\nu = \Omega/2\pi = 1.116$ kHz and the amplitude $f$ has been taken as the control parameter. The circuit exhibits a rich variety of bifurcations and chaos in its dynamics as the control parameter $f$ is increased from zero. The rich variety of bifurcations and chaos exhibited by the $MLCV$ circuit has been studied experimentally and numerically \cite{Thamilmaran2000,Thamilmaran2001}. An explicit analytical solution to the normalized state equations of the circuit has been presented \cite{Thamilmaran2001}. The circuit exhibits chaotic attractors at two values of the control paramter $f$. The attractor at $f=0.39$ is obtained through breakdown of a torus attractor while the chaotic attarctor at $f=0.411$ is obtained through period-doubling sequneces and leads to a reverse period doubling sequence indicating the antimonocity behavior. The chaotic attractors at $f=0.39$ and $f=0.411$ along with their correponding power spectra, indicating a broad band nature of frequencies is shown in Fig. \ref{fig:5}(a) and \ref{fig:5}(b) respectively. 

In the uncoupled state, the solutions to the state variables of the response system $x^{'},~y^{'}$ remains the same as that of the drive system except that the response operates with a different set of initial condition. The synchronization dynamics of the coupled circuits is studied by fixing the response system at two different chaotic states. In the first case, the response system is kept in the chaotic state obtained at $f_2=0.39$ while in the second case it is operated at the chaotic state obtained for $f_2=0.411$. \\
The analytical soulutions obtained in Section \ref{sec:3} can be used to generate two-parameter bifurcation diagrams to identify the nature of synchronization observed in the coupled system. Figures \ref{fig:6}(a) and \ref{fig:6}(b) shows the analytically obtained two-parameter bifurcation diagrams obtained in the $(\epsilon-f_1)$ plane for the response system operating at the chaotic states $f_2=0.39$ and $f_2=0.411$, respectively. The different synchronization regimes are color coded as follows: red-unsynchronized state (US), green-complete synchronization state (CS) and blue-phase synchronized state (PS). From  Figs. \ref{fig:6}(a) and \ref{fig:6}(b) we can infer that the coupled systems remain unsynchronized $(US)$ for smaller values of the coupling strength and gets completely synchronized for higher values. However, complete synchronization $(CS)$ of chaotic attractors are observed only when the drive and the response systems are operated in identical chaotic states. In Fig. \ref{fig:6}(a), we can observe that the coupled system exhibits CS for larger values of coupling strength when the response system is operated at the identical chaotic state of the drive i.e., at $f_1=0.39$. In the non-identical chaotic state i.e., $f_1=0.411$, phase synchronization $(PS)$ of the coupled systems is observed. Similarly, when the response system is operated in the chaotic state of $f_2 = 0.411$, CS is observed at the identical chaotic states $(f_1=0.411)$ and and PS at the non-identical chaotic state $(f_1=0.39)$. In this section we analyze the phenomenon of CS and PS observed in the coupled {\emph{variant of MLC circuits}} using the analytical solutions obtained in Section \ref{sec:3}.

Before we procced to the synchronization dynamics we present a linear stability analysis of the difference system observed in each of the piecewise-linear region. The eigenvalues $m_1$ and $m_2$ in the $D^{*}_0$ region are found to be a pair of complex conjugates for $\epsilon < 0.5682135$ while they are real and distinct for $\epsilon \ge 0.5682135$. Similarly, in the $D^{*}_{\pm 1}$ region, the eigenvalues $m_3$ and $m_4$ are found to be a pair of complex conjugates for $\epsilon < 0.0519135$ while they are real and distinct for $\epsilon \ge 0.0519135$. The transformation of the eigenvalues of the difference systems in the $D^{*}_0$, $D^{*}_{\pm 1}$ regions are summarized in Table \ref{tab:1}. Fig.\ref{fig:7}(a) and \ref{fig:7}(b) shows the bifurcation of the real eigenvalues of the Jacobian matrix given by Eq. \ref{eqn:14} and \ref{eqn:15} in the $D^{*}_{0}$ and $D^{*}_{\pm1}$ regions respectively as functions of the coupling parameter. The red and green lines show the two real roots while the blue line shows the real part of the complex conjugate roots. With the increase in coupling parameter, the fixed points in the $D^{*}_{0}$ and $D^{*}_{\pm1}$ regions transform into {\emph{stable nodes}} indicating the asymptotic convergence of trajectories towards the origin, in their corresponding regions of phase space, within the synchronization manifold. 

\begin{table}
\caption{Stability of fixed points in $D^{*}_{0}$ and $D^{*}_{\pm1}$ regions of the difference system. The fixed points transform into {\emph{stable nodes}} in all the three regions as the coupling paramter is increased.}
\begin{ruledtabular}
\begin{tabular}{c c c c c}
Region		&  Fixed Point		&	 $\epsilon$					&		Eigenvalues   		&	 Stability						\\ \hline
$D^{*}_{0}$	&	(0,0)			&        $\epsilon < 0.5682135$ 		&		Complex conjugates	& 	Unstable spiral					\\
			&				&        $\epsilon \ge 0.5682135$		&		Real and distinct		& 	Stable node					\\ \hline
$D^{*}_{\pm1}$	&	(0,0)		&	$\epsilon < 0.0519135$		&		Complex conjugates	& 	Stable spiral					\\
			&				&        $\epsilon \ge 0.0519135$		&		Real and distinct		& 	Stable node					\\ 
\end{tabular}
\end{ruledtabular}
\label{tab:1} 
\end{table}

\subsection{Complete Synchronization}
\label{sec:4a}
For studying somplete synchronization the amplitude of the response system is fixed at $f_2 = 0.39$. From Fig. \ref{fig:6}(a) we observe that complete synchronization is obtained when the drive and the response systems exist in identical chaotic states. The amplitude and frequency of the drive and the response sytems are fixed at $f_{1,2}=0.39$, $\omega_{1,2}=0.105$ while their initial conditions has been fixed at $x_0=-0.5,~y_0=0.1$ and $x^{'}_0 = 0.5,~y^{'}_0=0.11$, respectively. Owing to the difference in their initial conditions, the chaotic attractors of the drive and response become unsynchronized for $\epsilon = 0$. Fig. \ref{fig:8}(a) shows the unsynchronized state of the coupled system for the coupling parameter $\epsilon=0$, in the $(x-x{'})$ phase plane and their corresponding trajectory in the $x^{*}=x-x^{'}$ plane in Fig. \ref{fig:8}(b). As the value of the coupling parameter is increased, the trajectory of the response system begins to converge towards the trajectory of the drive and gets completely synchronized with the drive. Fig. \ref{fig:8}(c) shows the complete synchronization of the coupled systems in the $(x-x{'})$ phase plane for the value of the coupling parameter $\epsilon = 0.2$ and the corresponding trajectory in the $(x, x^{'})$ plane as in Fig. \ref{fig:8}(d). From the phase portraits and the time series plots obtained it could be inferred that for the coupling parameter taking the value $\epsilon=0.2$, the response system which is operating with a different set of initial condition, completely synchronizes with the drive. From the analytical results obtained we  conclude that each piecewise linear region of the response system is controlled by its counterpart in the drive system. The mechanism of synchronization could be understood from the behavior of the eigenvalues in the $D^{*}_{\pm1}$ region as a function of the coupling parameter. The transformation of the fixed point in the $D^{*}_{\pm1}$ region from a {\emph{stable focus}} to a {\emph{stable node}} for $\epsilon > 0.0519$ enables the trajectory of the response to mimic the trajectory of the drive. Hence the dynamics of the response is controlled by the drive and is oblidged to exist within the phase space of the drive. 

%

The stability of the synchronized state for the x-coupled systems given by Eqs. \ref{eqn:3} and \ref{eqn:9} is studied using the MSF approach. The state variables of the drive and the response system are represented as $(x, y, \theta)$ and $(x^{'}, y^{'}, \theta^{'})$, respectively. The MSF is generally the largest transverse Lyapunov exponent $(\lambda_{max})$ of the variational equation \cite{Pecora1998}. Figure \ref{fig:9} shows a plot of the MSF $(\lambda_{max})$ as a function of the coupling paramter $(\epsilon)$ for the $x \rightarrow x^{'}$ coupling of the state variables. The x-coupled system enters into a state of complete synchronization for $\epsilon > 0.003$, marked by the negative values of MSF. There exists a stable synchronized state in the region, $0.003 < \epsilon < 28.5$, within which the MSF is negative. For $\epsilon > 28.5$, the coupling between the systems become stronger and the synchronized state becomes unstable leading to positive values of the MSF. Hence the x-coupled system enters into the synchronized state even for smaller values of coupling strength and becomes unsynchronized for much higher values of the coupling strength.\\

The Lyapunov exponents of the attractor shown in Fig. \ref{fig:8}(a) are, $\lambda_1 = 0.0021,~\lambda_2 = -0.1262,~\lambda_3 = 0,~\lambda_4 = 0.0036,~\lambda_5 = -0.1268$ and $\lambda_6 = 0$. Hence, in the unsynchronized state we have two positive Lyapunov exponents indicating the chaotic state of the drive and response systems. The Lyapunov exponents of the coupled system for the coupling strength $\epsilon = 0.2$ are, $\lambda_1 = 0.0037,~\lambda_2 = -0.0702,~\lambda_3 = 0,~\lambda_4 = -0.1263,~\lambda_5 = -0.2559$ and $\lambda_6 = 0$. The negative values of the conditional Lyapunov exponents $\lambda_{4,5}$ confirms the synchronization of the response system with the drive. Fig. \ref{fig:10}(a) and \ref{fig:10}(b) shows the negative conditional Lyapunov exponents $\lambda_{4,5}$ and the {\emph{Kaplan-Yorke dimension}} $(D_{KY})$ as functions of the coupling parameter. A sudden fall in the $D_{KY}$ of the coupled system indicates the synchronization of the systems \cite{Steffen2013}.  

\subsection{Phase Synchronization}
\label{sec:4b}

In this section we discuss the phenomenon of phase synchronization obtained when the drive and the response systems are operated at non-identical chaotic states. For studying phase synchronization, we consider the case of the response sytem deing kept at the chaotic state $f_2=0.411$. From the two-parameter bifurcation diagram shown in Fig. \ref{fig:6}(b), we can observe that phase synchronization of the coupled systems takes place when $f_1=0.39$, for larger values of coupling strength. Hence, the coupled system is studied for the synchronization of their phases for different values of coupling strength as shown in Fig. \ref{fig:11}. The phases of the two systems which are initially unsynchronized for $\epsilon=0.01$ undergoes an imperfect synchronization of phases for $\epsilon=0.123$ and are perfectly synchronized for $\epsilon=0.131$. The phase portraits of the attarctors indicating the unsynchronized, imperfectly and perfectly phase synchronized states in the $(x-x^{'})$ phase planes is shown in Fig. \ref{fig:12}(a). The corresponding time-series plot of the signals $x$ and $x^{'}$ shown in Fig. \ref{fig:12}(b) clearly indicates the mismatch in the amplitues of the drive and response signals inspite of the synchronization of their phases for all the values of the coupling strengths. The negative values of the conditional lyapunov exponents of the response system $(\lambda_{4,5})$ and the sudden fall in the $D_{KY}$ of the coupled system shown in Fig. \ref{fig:13}(a) and \ref{fig:13}(b), confirms the synchronization of the coupled system.

\section{Synchronization dynamics of Forecd parallel LCR circuit with simplified nonlinear element}
\label{sec:5}

In this section we discuss the synchronization dynamics observed in coupled parallel LCR circuits with the {\emph{simplified nonlinear element}} being the nonlinear element. The circuit parameters have been chaosen as   $C=13.13$ nF, $L=163.6$ mH, $R=2.05~k \Omega$. The negative slopes of the inner, outer regions and the breakpoints in the $(v-i)$ characteristic curve of the piecewise-linear element shown in Fig. \ref{fig:1}(b) are given as $G_{a} = -0.56$ mS, $G_{b} = +2.5$ mS and $B_{p} = \pm~3.8$ V, respectively. The rescaled circuit parameters take the values as $\beta = 0.2592,~a=-1.148,~b=5.125$. The frequency of the external periodic force has been fixed at $\nu = \Omega/2\pi = 1.678$ kHz and the amplitude $f$ has been taken as the control parameter. The circuit exhibits a rich spectrum of dynamical phenomenon as the control parameter $f$ is increased from zero. The circuit exhibits chaotic attractors at two values of the control paramter $f=0.695$ and $f=0.855$, respectively. The two chaotic attarctors along with their correponding power spectra is shown in Fig. \ref{fig:14}(a) and \ref{fig:14}(b) respectively.

As with the case of the {\emph{Variant of MLC circuit}}, this circuit has also been studied for the synchronization of identical and non-identical chaotic attractors. Two parameter bifurcation diagrams in the $(\epsilon - f_1)$ plane for the response systems being fixed at the chaotic states $f_2=0.695$ and $f_2=0.855$,respectively, are as shown in Fig. \ref{fig:15}(a) and \ref{fig:15}(b), respectively. Complete synchronization (CS) is observed when the drive and response systems have been kept at identical chaotic states while phase synchronization (PS) is observed when the two systems are kept at non-identical chaotic states. In the following discussion, we present the mechanism of CS and PS observed in the coupled system. In the first case, the drive and the respose systems have been kept in the same chaotic stae i.e. $f_{1,2}=0.695$, while they have been operated with two different sets of initial conditions. In the second case, in addition to the different initial conditions, the drive is kept at the chaotic state at $f_1=0.695$ while the response system at $f_2=0.855$. The linear stability analysis on the eigenvalues of the difference system shows a bifuraction of the eigenvalues in the $D^{*}_0$ region as function of the coupling parameter. The eigenvalues $m_1$ and $m_2$ in the $D^{*}_0$ region are found to be a pair of complex conjugates for $\epsilon < 1.1663$ while they are real and distinct for $\epsilon \ge 1.1663$. In the $D^{*}_{\pm 1}$ region, the eigenvalues $m_3$ and $m_4$ are found to be real, negative and distinct for all values of the coupling strength. The bifurcation of the real eigenvalues in the $D_0$ region is shown in Fig. \ref{fig:16}. The red and green lines show the two real roots while the blue line shows the real part of the complex conjugate roots. With the increase in coupling parameter, the fixed point of the $D^{*}_{0}$ region transform into {\emph{stable nodes}}.

\subsection{Complete Synchronization}   

In this case, the drive and the response systems are kept in identical chaotic states obtained at $f_{1,2}=0.695$. However, they evolve from two different initial conditions given by $x_0=-0.5,~y_0=0.1$ and $x^{'}_0 = 0.5,~y^{'}_0=0.11$, respectively. Owing to the difference in the initial conditions, the chaotic attractors are initially unsynchronized as shown in Fig. \ref{fig:17}(a), for $\epsilon=0$. Hence the uncoupled system must have two positive lyapunov exponents given by $\lambda_1 = 0.0395,~\lambda_2 = -2.99062,~\lambda_3 = 0,~\lambda_4 = 0.03277,~\lambda_5 = -2.96527$ and $\lambda_6 = 0$, as expected. Figure \ref{fig:17}b(i) shows the time-series of the state variables $x$ (blue line) and $x^{'}$ (magenta line) corresponding to the attractor shown in Fig. \ref{fig:17}a(i). As the coupling strength is increased in the range $0.007 \le \epsilon \le 0.014$, an enhancement in the magnitude of the amplitude of response signals has been noticed. This is observed by an increase in the phase space of the attractor as shown in  Fig. \ref{fig:17}(b). Further, an analysis on the lyapunov exponents reveals the existance of two positive lyapunov exponents for the coupled system and hence confirming the hyperchaotic behavior of the atrtractor shown in  Fig. \ref{fig:17}(b). The lyapunov exponents for the coupling strength $\epsilon=0.01$ has been given by $\lambda_1 = 0.03996,~\lambda_2 = 0.00276,~\lambda_3 = 0,~\lambda_4 = -2.83489,~\lambda_5 = -3.02702$ and $\lambda_6 = 0$. The time-series plot shown in Fig. \ref{fig:17}b(ii) indicates the amplitude enhanced response signal $x^{'}$ (magenta line) with that of the drive signal $x$ (blue line). With further increase in the coupling strength, the response system completely synchronizes with the drive. Figure \ref{fig:17}a(iii) shows the complete synchronized state of the coupled system in the $(x-x^{'})$ phase plane for $\epsilon=0.181$. The time-series of the state variables $x,x^{'}$ Fig. \ref{fig:17}b(iii) clearly shows the complete synchronization of the coupled system. Fig. \ref{fig:18} shows the two largest lyapunov exponents $\lambda_1, \lambda_2$ of the couple system. The positive values of the $\lambda_1$ and $\lambda_2$ in the region $0 \le \epsilon \le 0.014$ confirms the hyperchaotic nature of the attractor shown in Fig. \ref{fig:17}(b). The stability of the completely synchronized state is indicated by the negative values of the MSF $(\lambda_{max})$ in the region $0.0512 \le \epsilon \le 26.8719$, as shown in Fig. \ref{fig:19}. The negative values of the conditional lyapunov exponents $(\lambda_{4,5})$ and the $D_{KY}$ are shown in Fig. \ref{fig:20}(a) and \ref{fig:20}(b), respectively. The $D_{KY}$ showing continuous variations in the dimension of the system for $\epsilon \le 0.045$ shows a sudden fall for $\epsilon>0.045$ indicating the synchronization of the coupled systems. The phenomenon of complete synchronization has been evidenced in coupled identical chaotic attarctors through an emergence of hyperchaoticity in its dynamics.

\subsection{Phase Synchronization} 

In this case, the drive and the response systems are operated at non-identical chaotic states obtained at $f_{1}=0.695$ and $f_2=0.855$, respectively. Similar to the case discussed with the {\emph{variant of MLC circuit}}, the non-identical chaotic circuits synchronizes in their phases as the coupling strength is increased. Fig.~\ref{fig:21} shows the phase difference $(\phi_1 - \phi_2)$ of the drive and the response systems for different values of the coupling strength. The phases of the two system which are initially unsynchronized for $\epsilon=0.001$ undergoes an imperfect synchronization of phases for $\epsilon=0.149$ and has been perfectly synchronized for $\epsilon=0.17$. The phase portraits of the attarctors indicating the unsynchronized, imperfectly and perfectly phase synchronized states in the $(x-x^{'})$ phase planes is shown in Fig. \ref{fig:22}(a). The corresponding time-series plot of the signals $x$ and $x^{'}$ shown in Fig. \ref{fig:22}(b) clearly indicates the mismatch in the amplitues of the drive and response signals inspite of the synchronization of their phases for all the values of the coupling strengths. The negative values of the conditional lyapunov exponents of the response system and the sudden fall in the $D_{KY}$ of the coupled system shown in Fig. \ref{fig:23}(a) and \ref{fig:23}(b), confirms the synchronization of the coupled system.

\section{Conclusion}

We have presented in this paper, a generalized explicit analytical solution for coupeld forced parallel LCR circuits with three-segmented piecewise linear nonlinear element. Because the chaotic circuits studied here differ only by their nonlinear elements, the mathematical formulation of their normalized state equations remains the same. Hence, a generalized analytical solution is arrived for the state equations and are used to study the synchronization dynamics of each system. The circuit with the {\emph{Chua's diode}} as the nonlinear element presents complete and phase synchronized states for coupling of identical and non-identical chaotic regimes. However, the circuits with the {\emph{simplified nonlinear element}} exhibits complete synchronization through the evolution of a hyperchaotic attarctor when coupled in identical chaotic regime. Further, amplitude enhancement is observed in its dynamics along the path to complete and phase synchronization. The entire dynamics of the two circuit systems are observed through two-parameter bifurcation diagrams obtained from the explicit analytical solutions. The mechanism of complete synchronization is identified through linear stability analysis of the difference system in each of the coupled regions. The bifurcation of the eigenvalues as functions of the coupling parameter in each of the regions reveals the onset of synchronization. The results obtained from the analytical solutions have been substantiated by solid numerical results. The analytical method presented in this study is simple as it involves only the coupling of similar regions of the drive and response and finding the solutions to each region of operation. This method has been checked earlier with yet another simple chaotic circuit namely, the {\emph{Murali-Lakshmanan-Chua}} circuit \cite{Sivaganesh2015}. In this paper we have confirmed the reliability of this analytical method and its solution to unidirectionally coupled picewise linear systems, by presenting two-parameter bifurcation diagrams. Hence we suggest that this method could be applied for an effective study of synchronization and its mechanism in unidirectionally coupled piecewise linear systems.

\pagebreak

\begin{figure}
\resizebox{1\textwidth}{!}{%
  \includegraphics{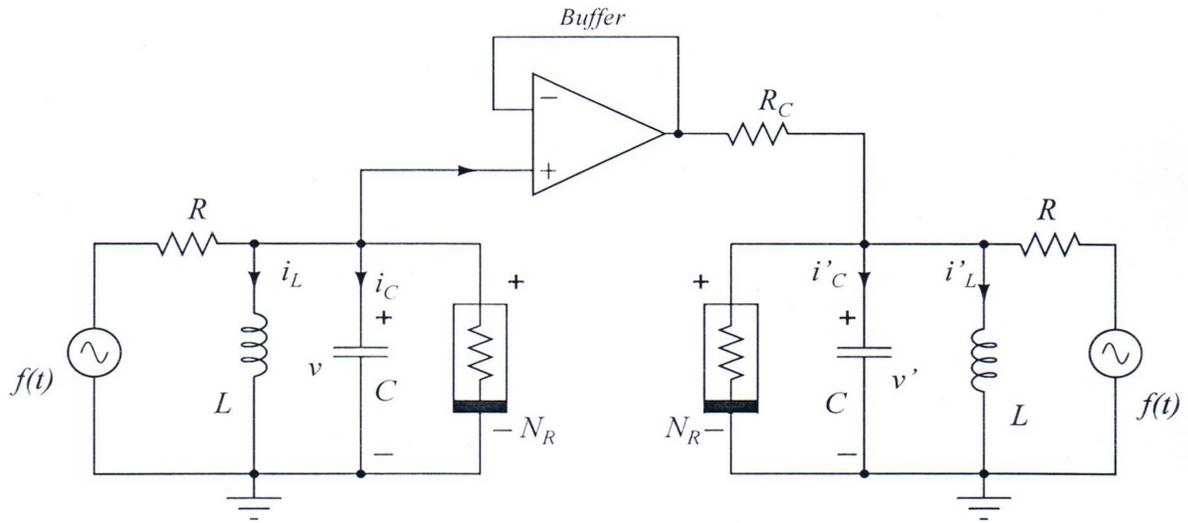}
}
\caption{Schematic circuit realization of unidirectionally coupled forced parallel LCR circuits with nonlinear elements $N_R$.}
\label{fig:4}       
\end{figure}

\end{document}